\documentclass[draft,tightenlines,nofootinbib,preprint,aps,eqsecnum,amsmath,amssymb]{revtex4}

\newcommand{\beq}{\begin{equation}}
\newcommand{\eeq}{\end{equation}}
\newcommand{\bea}{\begin{eqnarray}}
\newcommand{\eea}{\end{eqnarray}}

\newcommand{\sgn}{\mbox{\boldmath $\epsilon$}}

\begin{document}

\baselineskip 18pt

\title{Relativistic Atomic Physics: from Atomic Clock
Synchronization towards Relativistic Entanglement.}

\medskip

\author{Luca Lusanna}

\affiliation{ Sezione INFN di Firenze\\ Polo Scientifico\\ Via Sansone 1\\
50019 Sesto Fiorentino (FI), Italy\\ E-mail: lusanna@fi.infn.it}

\begin{abstract}

A review is given of the implications of the absence of an intrinsic
notion of instantaneous 3-space, so that a clock synchronization
convention has to be introduced, for relativistic theories.

\bigskip\bigskip\bigskip\bigskip\bigskip\bigskip\bigskip\bigskip

Talk given at the 7th Friedmann International Seminar, Joao Pessoa
(Brasil), June 29 - July 5, 2008

\end{abstract}

\maketitle

\vfill\eject

The recent developments in the construction of microwave and optical
clocks in atomic physics are opening a new era in clock
synchronization \cite{1}. The proposed ACES mission of ESA, if
accepted, will make possible a measurement of the gravitational
redshift of the Earth from the two-way link among a microwave clock
(PHARAO) on the Space Station and similar clocks on the ground: the
proposed microwave link should make possible the control of effects
on the scale of 5 picoseconds. This will be a test of post-Newtonian
gravity in the framework of Einstein's geometrical view of
gravitation: the redshift is a measure of the $1/c^2$ deviation of
post-Newtonian null geodesics from Minkowski ones. This is going to
create problems to relativistic metrology (the standard of time will
have to be put in space to avoid the local variations of the
geopotential) and will open the possibility of relativistic geodesy
for formulating a theory of heights over the reference geoid.
\bigskip

The problem of clock synchronization is equivalent to the problem of
the definition of an instantaneous 3-space (all its points are
synchronous), which in turn is a prerequisite for the definition of
a well-posed Cauchy problem for field equations like the Maxwell
ones, i.e. for the predictability of the future.
\bigskip

In Galilei space-time both Newtonian time and the Euclidean 3-space
(with the associated notion of spatial distance) are {\it absolute}
so that the problem of clock synchronization does not exist in
either inertial or non-inertial frames. The inertial ones, connected
by Galilei transformations, are an ideal limit selected by Newton
law of inertia and by the Galilei relativity principle: in them
Newton's equations are invariant in form. The apparent forces of
non-inertial frames are proportional to the inertial mass, which in
turn is equal to the gravitational mass (the Galilei equivalence
principle). Non-relativistic quantum mechanics, with its
foundational problems, and the theory of entanglement are formulated
in this framework where Maxwell equations do not exist. The photons
in the discussions about entanglement and teleportation are only
states with two polarizations in a two-dimensional Hilbert space:
their carrier cannot be a ray of light in the eikonal approximation
moving along a null geodesic, because such null path does not exist
in Galilei space-time. The existing inclusion of electro-magnetism
at the order $1/c$ made by atomic physics destroys the Galilei group
and does allow a consistent definition of the Poincare' one. It is
enough for experiments on the Earth, but not for going to space like
in the ACES mission.
\bigskip

In special relativity the only intrinsic structure available to a
time-like observer in Minkowski space-time is the conformal one (the
light-cone): it is the locus of the incoming or outgoing rays of
light. There is no notion of simultaneity, of instantaneous 3-space,
of spatial distance. The light postulates say that the two-way (or
round-trip; only one clock is involved) velocity of light is a)
isotropic and b) constant (a standard constant $c$ replaces the
standard of length in relativistic metrology). The one-way velocity
of light between two observers depends on how their clocks are
synchronized (in general is not isotropic and point-dependent). The
ideal inertial frames centered on inertial observers, connected by
Poincare' transformations and with the physical laws invariant in
form due to the relativity principle, can be identified with
Einstein's convention for clock synchronization: an inertial
observer A send a ray of light at $x^o_i$ towards the observer B;
the ray is reflected towards A at a point P of B world-line and then
reabsorbed by A at $x^o_f$; by convention P is synchronous with the
mid-point between emission and absorption on A world-line, i.e.
$x^o_P = x^o_i + {1\over 2}\, (x^o_f - x^o_i)$. This convention
selects the Euclidean instantaneous 3-spaces $x^o = ct = const.$ of
the inertial frames centered on A. Only in this case the one-way
velocity of light between A and B coincides with the two-way one,
$c$. As a consequence in relativistic metrology the Euclidean
spatial length between A and B is defined as ${1\over 2}\, c\,
(x^o_f - x^o_i)$.\bigskip

However if the observer A is accelerated the convention breaks down.
This is due to the fact that if we know {\it only} the world-line of
the accelerated observer (the 1+3 point of view) the only way for
defining instantaneous 3-spaces is to identify them with the
Euclidean tangent planes orthogonal to the 4-velocity of the
observer (the local rest frames): these planes intersect each other
at a distance from A world-line of the order of the acceleration
lengths of A \cite{2} ($l = c^2/a$ for linear acceleration $a$ and
$l = c/\omega$ for rotational angular velocity $\omega$). Therefore
all the accelerated frames, centered on accelerated observers, based
either on Fermi coordinates or on rotating ones will develop {\it
coordinate singularities}, so that their instantaneous 3-spaces
cannot be used for a well-posed Cauchy problem for Maxwell
equations. For the rotating disk the coordinate singularity appears
at a distance $R$ from the rotation axis where $\omega\, R = c$ (the
so-called "horizon problem"). According to the locality hypothesis
for the theory of measurements \cite{2} an accelerated observer is
identified with a succession of instantaneously comoving inertial
observers. See Refs\cite{3} for a rich bibliography on these topics.
\bigskip

The way out from these problems is the 3+1 point of view \cite{4},
in which we assign: a) the world-line of an arbitrary time-like
observer; b) an admissible 3+1 splitting of Minkowski space-time,
namely a nice foliation with space-like instantaneous 3-spaces (i.e.
a clock synchronization convention). This allows to define a {\it
global non-inertial frame} centered on the observer and to use
observer-dependent Lorentz-scalar {\it radar 4-coordinates}
$\sigma^A = (\tau ;\sigma^r)$, where $\tau$ is a monotonically
increasing function of the proper time of the observer and
$\sigma^r$ are curvilinear 3-coordinates on the 3-space
$\Sigma_{\tau}$ having the observer as origin. If $x^{\mu} \mapsto
\sigma^A(x)$ is the coordinate transformation from the inertial
Cartesian 4-coordinates $x^{\mu}$ to radar coordinates, its inverse
$\sigma^A \mapsto x^{\mu} = z^{\mu}(\tau ,\sigma^r)$ defines the
embedding functions $z^{\mu}(\tau ,\sigma^r)$ describing the
3-spaces $\Sigma_{\tau}$ as embedded 3-manifold into Minkowski
space-time. The induced 4-metric on $\Sigma_{\tau}$ is the following
functional of the embedding $g_{AB}(\tau ,\sigma^r) = [z^{\mu}_A\,
\eta_{\mu\nu}\, z^{\nu}_B](\tau ,\sigma^r)$, where $z^{\mu}_A =
\partial\, z^{\mu}/\partial\, \sigma^A$ and $\eta_{\mu\nu} = \sgn\,
(+---)$ is the flat metric ($\sgn = \pm 1$ according to either the
particle physics $\sgn = 1$ or the general relativity $\sgn = - 1$
convention). While the 4-vectors $z^{\mu}_r(\tau ,\sigma^u)$ are
tangent to $\Sigma_{\tau}$, so that the unit normal $l^{\mu}(\tau
,\sigma^u)$ is proportional to $\epsilon^{\mu}{}_{\alpha
\beta\gamma}\, [z^{\alpha}_1\, z^{\beta}_2\, z^{\gamma}_3](\tau
,\sigma^u)$, we have $z^{\mu}_{\tau}(\tau ,\sigma^r) = [N\, l^{\mu}
+ N^r\, z^{\mu}_r](\tau ,\sigma^r)$ ($N(\tau ,\sigma^r) = \sgn\,
[z^{\mu}_{\tau}\, l_{\mu}](\tau ,\sigma^r)$ and $N_r(\tau ,\sigma^r)
= - \sgn\, g_{\tau r}(\tau ,\sigma^r)$ are the lapse and shift
functions).\medskip

Let us remark that both the 1+3 and the 3+1 points of view are {\it
non factual}; in both of them one must know an entire world-line
from $\tau = - \infty$ to $\tau = + \infty$ and in the 3+1 one also
a whole instantaneous 3-space.

\medskip

The foliation is nice and admissible if it satisfies the conditions:
1) $N(\tau ,\sigma^r) > 0$ in every point of $\Sigma_{\tau}$ (the
3-spaces never intersect); 2) $\sgn\, g_{\tau\tau}(\tau ,\sigma^r) >
0$, so to avoid the horizon problem of the rotating disk, and with
the positive-definite 3-metric $h_{rs}(\tau ,\sigma^u) = - \sgn\,
g_{rs}(\tau ,\sigma^u)$ having three positive eigenvalues (these are
the M$\o$ller conditions \cite{5}); 3) all the 3-spaces
$\Sigma_{\tau}$ must tend to the same space-like hyper-plane at
spatial infinity (so that there are always asymptotic inertial
observers to be identified with the fixed stars). As a consequence
{\it rigid} rotations are forbidden in relativistic theories: see
Refs.\cite{3} for the simplest example of admissible 3+1 splitting
with differential rotations. Each nice foliation has two associated
congruences of time-like observers: a) the Eulerian ones having the
unit normal $l^{\mu}(\tau ,\sigma^r)$ to $\Sigma_{\tau}$ as
4-velocity; b) the rotating observers having
$[z^{\mu}_{\tau}/\sqrt{\sgn\, g_{\tau\tau}}](\tau ,\sigma^r)$ as
4-velocity (this congruence is not surface-forming like the ones
simulating the rotating disks).\medskip

The 4-metric $g_{AB}(\tau ,\vec \sigma )$ on $\Sigma_{\tau}$ has the
components $\sgn\, g_{\tau\tau} = N^2 - N_r\, N^r$, $- \sgn\,
g_{\tau r} = N_r = h_{rs}\, N^s$, $h_{rs} = - \sgn\, g_{rs} =
\sum_{a=1}^3\, e_{(a)r}\, e_{(a)s} = \gamma^{1/3}\, \sum_{a=1}^3\,
e^{2\, \sum_{\bar b =1}^2\, \gamma_{\bar ba}\, R_{\bar b}}\,
V_{ra}(\theta^i)\, V_{sa}(\theta^i)$), where $e_{(a)r}(\tau
,\sigma^u)$ are cotriads on $\Sigma_{\tau}$, $\gamma (\tau
,\sigma^r) = det\, h_{rs}(\tau ,\sigma^r)$ is the 3-volume element
on $\Sigma_{\tau}$, $\lambda_a(\tau ,\sigma^r) = [\gamma^{1/6}\,
e^{\sum_{\bar b =1}^2\, \gamma_{\bar ba}\, R_{\bar b}}](\tau
,\sigma^r)$ are the positive eigenvalues of the 3-metric
($\gamma_{\bar aa}$ are suitable numerical constants) and
$V(\theta^i(\tau ,\sigma^r))$ are diagonalizing rotation matrices
depending on three Euler angles. The components $g_{AB}$ or the
quantities $N$, $N_r$, $\gamma$, $R_{\bar a}$, $\theta^i$, play the
role of the {\it inertial potentials} generating the relativistic
apparent forces in the non-inertial frame. It can be shown \cite{6}
that the Newtonian inertial potentials are hidden in the functions
$N$, $N_r$ and $\theta^i$.\medskip

Let us remark that in the ADM Hamiltonian formulation of general
relativity in the York canonical basis of Ref.\cite{7}: a) the
quantities $R_{\bar a}(\tau ,\sigma^r)$, $\bar a =1,2$, become the
physical {\it tidal} degrees of freedom of the gravitational field
(the polarizations of the gravitational waves in the linearized
theory); b) the 3-volume element $\gamma (\tau ,\sigma^r)$ is
determined by the super-Hamiltonian constraint (the Lichnerowicz
equation) in terms of the other variables; c) there is an extra
inertial potential determining the allowed clock synchronization
conventions, i.e. the trace $K(\tau ,\sigma^r)$ of the extrinsic
curvature of the non-Euclidean 3-space $\Sigma_{\tau}$, which is a
functional of $g_{AB}$ in special relativity and has {\it no
Newtonian counterpart}. These results hold in a special class of
globally hyperbolic, asymptotically Minkowskian at spatial infinity,
topologically trivial space-time without super-translations so that
the asymptotic symmetries are reduced to the ADM Poincare' group as
shown in Refs.\cite{8} and the allowed 3+1 splittings of the
space-time allow to define the same type of global non-inertial
frames as in special relativity. However now the equivalence
principle says that global inertial frames do not exist, so that the
kinematical Poincare' group is replaced by the spatio-temporal
diffeonorphism group and the relativity principle with the principle
of general covariance (invariance in form of physical laws). Since
the absence of super-translations implies that the instantaneous
3-spaces are asymptotically orthogonal to the ADM 4-momentum, these
3-spaces are non-inertial rest frames of the 3-universe and admit
asymptotic inertial observers (the fixed stars). Moreover, if we
switch down the Newton constant, we get the description of the
matter present in these space-times in the non-inertial rest frames
of Minkowski space-time (deparametrization of general relativity)
with the ADM Poincare' group collapsing in the Poincare' group of
particle physics. However, in general relativity every solution of
Einstein equations {\it dynamically} selects its preferred
instantaneous 3-spaces (modulo coordinate transformations) \cite{9}:
since the whole chrono-geometrical structure, described by the
4-metric and the associated line element, is now {\it dynamical},
also the clock synchronization convention acquire a dynamical
character. The gravitational field, i.e. the 4-metric, is not only
the potential of the gravitational interaction but it also teaches
relativistic causality to the other fields (it says to each massless
particle which are the allowed trajectories in each point). This
geometrical property is lost when the 4-metric is split in a
background plus a perturbation (like in quantum field theory and
string theory for being able to define a Fock space), since the
chrono -geometrical structure is frozen to the one of the
background; in Refs.\cite{7,8,9} such a splitting is never done,
since there an {\it asymptotic} Minkowskian background.

\bigskip

Let us come back to special relativity and let consider any isolated
system (particles, strings, fields, fluids) admitting a Lagrangian
description allowing, through the coupling to an external
gravitational field, the determination of the matter energy-momentum
tensor and of the ten conserved Poincare' generators $P^{\mu}$ and
$J^{\mu\nu}$ (assumed finite) of every configuration of the system.
Let us replace the external gravitational 4-metric in the coupled
Lagrangian with the 4-metric $g_{AB}(\tau ,\sigma^r)$ of an
admissible 3+1 splitting of Minkowski space-time and let us replace
the matter fields with new ones knowing the instantaneous 3-spaces
$\Sigma_{\tau}$. For instance a Klein-Gordon field $\tilde \phi (x)$
will be replaced with $\phi(\tau ,\sigma^r) = \tilde \phi (z(\tau
,\sigma^r))$; the same for every other field. Instead for a
relativistic particle with world-line $x^{\mu}(\tau )$ we must make
a choice of its energy sign and it will be described by
3-coordinates $\eta^r(\tau )$ defined by the intersection of the
world-line with $\Sigma_{\tau}$: $x^{\mu}(\tau ) = z^{\mu}(\tau
,\eta^r(\tau ))$.

In this way we get a Lagrangian depending on the given matter and on
the embedding $z^{\mu}(\tau ,\sigma^r)$ and this formulation has
been called {\it parametrized Minkowski theories} \cite{10},
\cite{3,4}. These theories are invariant under frame-preserving
diffeomorphisms (see Ref.\cite{11} for their first identification as
the subgroup of space-time diffeomorphism of general relativity
relevant for non-inertial frames), so that there are four
first-class constraints (an analogue of the super-Hamiltonian and
super-momentum constraints of canonical gravity) implying that the
embeddings $z^{\mu}(\tau ,\sigma^r)$ are {\it gauge variables}. As a
consequence, all the admissible non-inertial frames are gauge
equivalent, namely physics does {\it not} depend on the clock
synchronization convention: only the appearances of phenomena change
by changing the notion of instantaneous 3-space.

\bigskip

A particular case of this description is the {\it rest-frame instant
form of dynamics for isolated systems} \cite{10}, \cite{3,4} which
is done in the intrinsic inertial rest frame of their
configurations: the instantaneous 3-spaces, named Wigner 3-space due
to the fact that the 3-vectors inside them are Wigner spin-1
3-vectors, are orthogonal to the conserved 4-momentum of the
configuration (in Ref.\cite{6} there will be the extension to
non-inertial rest frames like the ones in the formulation of
canonical gravity previously quoted). In this rest frames there are
only three notions of collective variables, which can be built by
using {\it only} the Poincare' generators (they are {\it non-local}
quantities knowing the whole $\Sigma_{\tau}$) \cite{12}: The
canonical non-covariant Newton-Wigner center of mass (or center of
spin), the non-canonical covariant Fokker-Pryce center of inertia
and the non-canonical non-covariant M$\o$ller center of energy. All
of them tend to the Newtonian center of mass in the non-relativistic
limit. See Ref.\cite{4} for the M$\o$ller non-covariance world-tube
around the Fokker-Pryce 4-vector identified by these collective
variables. As shown in Refs.\cite{12,13,14} these three variables
can be expressed as known functions of the rest time $\tau$, of the
canonically conjugate Jacobi data (frozen Cauchy data) $\vec z =
Mc\, {\vec x}_{NW}(0)$ (${\vec x}_{NW}(\tau )$ is the standard
Newton-Wigner 3-position) and $\vec h = \vec P/Mc$, of the invariant
mass $Mc = \sqrt{\sgn\, P^2}$ of the system and of its rest spin
${\vec {\bar S}}$. It is convenient to center the inertial rest
frame on the Fokker-Pryce inertial observer.

As a consequence, every isolated system (i.e. a closed universe) can
be visualized as a decoupled non-covariant collective (non-local)
pseudo-particle described by the frozen Jacobi data $\vec z$, $\vec
h$ carrying a {\it pole-dipole structure}, namely the invariant mass
and the rest spin of the system, and with an associated external
realization realization of the Poincare' group. This structure
implements old ideas of Ref.\cite{15}. The universal breaking of
Lorentz covariance is connected to this decoupled non-local
collective variable and is irrelevant because all the dynamics of
the isolated system leaves inside the Wigner 3-spaces and is
Wigner-covariant. It turns out \cite{14} that there are three pairs
of second class (interaction-dependent) constraints eliminating the
internal 3-center of mass and its conjugate momentum inside the
Wigner 3-spaces: this avoids a double counting of the collective
variables and allows to re-express the dynamics only in terms of
internal Wigner-covariant relative variables. In the case of
relativistic particles the reconstruction of their world-lines
requires a complex interaction-dependent procedure delineated in
Ref.\cite{13}. See Ref.\cite{14} for the comparison with the other
formulations of relativistic mechanics developed for the study of
the problem of {\it relativistic bound states}.

\bigskip

In this framework it has been possible to obtain a relativistic
formulation of the classical background of atomic physics,
considered as an effective theory of positive-energy scalar (or
spinning) particles with mutual Coulomb interaction plus the
transverse electro-magnetic field of the radiation gauge valid for
energies below the threshold of pair production. As shown in
Refs.\cite{16} and \cite{14} (in Ref.\cite{17} there will be the
elimination of the internal 3-center of mass for this system), this
has been possible by considering Grassmann-valued electric charges
for the particles ($Q_i^2 = 0$, $Q_i\, Q_j = Q_j\, Q_i \not= 0$ for
$i \not= j$). It allows a) to make an ultraviolet regularization of
Coulomb self-energies; b) to make an infrared regularization
eliminating the photon emission; c) to express the Lienard-Wiechert
potentials only in terms of the 3-coordinates $\eta^r_i(\tau )$ and
the conjugate 3-momenta $\kappa_{ir}(\tau)$ in a way independent
from the used (retarded, advanced,..) Green function. All this
amount to reformulate the dynamics of the one-photon exchange as a
Cauchy problem with well defined potentials. Moreover there is a
canonical transformation \cite{14} sending the above system in a
transverse radiation field (in- or out-fields) decoupled, in the
global rest frame, from Coulomb-dressed particles with a mutual
interaction described by the sum of the Coulomb potential plus the
Darwin potential. Therefore for the first time we are able to obtain
results, previously derived from instantaneous approximations to the
Bethe-Salpeter equation for the description of relativistic bound
states (see the bibliography of Ref.\cite{16}), starting from the
classical theory. Moreover, for the first time, at least at the
classical level, we have been able to avoid the Haag theorem
according to which the interaction picture does not exist in
QFT.\bigskip

Let us now consider the quantum theory.\medskip

In refs.\cite{18} there is the quantization of positive-energy free
scalar and spinning particles in a family of non-inertial frames of
Minkowski space-time where the instantaneous 3-spaces are space-like
hyper-planes. We take the point of view {\it not to quantize the
inertial effects} (the appearances of phenomena): the embedding
$z^{\mu}(\tau ,\sigma^r)$ remains a c-number and we get results
compatible with atomic spectra. Instead the problem of the
reformulation of particle physics in non-inertial frames is unsolved
due to the no-go theorem of Ref.\cite{19} showing the existence of
obstructions to the unitary evolution of a massive Klein-Gordon
field between two space-like surfaces of Minkowski space-time. This
problem has to be reformulated as the search of the class of
admissible 3+1 splittings of Minkowski space-time admitting unitary
evolution after quantization: this would allow to check whether the
hypothesis of non-quantized inertial effects is valid also in field
theory (it will be a crucial point for quantum gravity!).

\bigskip

In Galilei space-time non-relativistic quantum mechanics, where all
the main results about entanglement are formulated, describes a
composite system with two (or more) subsystems with a Hilbert space
which is the tensor product of the Hilbert spaces of the subsystems:
$H = H_1 \otimes H_2$. This type of spatial separability is named
{\it the zeroth postulate} of quantum mechanics. However, when the
two subsystems are mutually interacting, one makes a unitary
transformation to the tensor product of the Hilbert space $H_{com}$
describing the decoupled Newtonian center of mass of the two
subsystems and of the Hilbert space $H_{rel}$ of relative variables:
$H = H_1 \otimes H_2 = H_{com} \otimes H_{rel}$. This allows to use
the method of separation of variables to split the Schroedinger
equation in two equations: one for the free motion of the center of
mass and another, containing the interactions, for the relative
variables (this equation describes both the bound and scattering
states). A final unitary transformation of the Hamilton-Jacobi type
allows to replace $H_{com}$ with $H_{com, HJ}$, the Hilbert space in
which the decoupled center of mass is frozen and described by
non-evolving Jacobi data. Therefore we have $H = H_1 \otimes H_2 =
H_{com} \otimes H_{rel} = H_{com, HJ} \otimes H_{rel}$.\medskip

While at the non-relativistic level these three descriptions are
unitary equivalent, this no more true in relativistic quantum
mechanics, the effective quantum theory for the description of atoms
as relativistic bound states of particles interacting through
action-at-a-distance potentials deduced from quantum field theory
(for instance the Coulomb plus Darwin potential). Once relativistic
quantum mechanics is under control, we can extend it to relativistic
atomic physics by quantizing also the transverse electro-magnetic
field in the radiation gauge.\medskip

As it will be shown in Ref.\cite{20}, the non-local and
non-covariant properties of the decoupled relativistic center of
mass, described by the frozen Jacobi data $\vec z$ and $\vec h$,
imply that the only consistent relativistic quantization is based on
the Hilbert space $H = H_{com, HJ} \otimes H_{rel}$. We have $H
\not= H_1 \otimes H_2$, because, already in the non-interacting
case, in the tensor product of two quantum Klein-Gordon fields,
$\phi_1(x_1)$ and $\phi_2(x_2)$,  most of the states correspond to
configurations in Minkowski space-time in which one particle may be
present in the absolute future of the other particle. This is due to
the fact that the two times $x^o_1$ and $x^o_2$ are totally
uncorrelated, or in other words there is no notion of instantaneous
3-space (clock synchronization convention). Also the scalar products
in the two formulations are completely different as shown in
Ref.\cite{21}. In S-matrix theory this problem is eliminated by
avoiding the interpolating states at finite (the problem of the Haag
theorem) and going the the asymptotic (in the times $x^o_i$) limit
of the free in- and out- states. However in atomic physics we need
interpolating states, and not S-matrix, to describe a laser beam
resonating in a cavity and intersected by a beam of atoms!\medskip

We have also $H \not= H_{com} \otimes H_{rel}$, because if instead
of $\vec z = Mc\, {\vec x}_{NW}(0)$ we use the evolving (non-local
and non-covariant) Newton-Wigner position operator ${\vec
x}_{NW}(\tau )$, then we get a violation of relativistic causality
because the center-of-mass wave packets spread instantaneously as
shown by the Hegerfeldt theorem \cite{22}.\medskip

Therefore the only consistent Hilbert space is $H = H_{com, HJ}
\otimes H_{rel}$, whose non-relativistic limit is the corresponding
Newtonian Hilbert space, corresponding to the quantization of the
inertial rest-frame instant form and englobing the notion of
instantaneous Wigner 3-spaces. The main complication is the
definition of $H_{rel}$, because we must take into account the three
pairs of (interaction-dependent) second-class constraints
eliminating the internal 3-center of mass inside the Wigner
3-spaces. When we are not able to make the elimination at the
classical level and formulate the dynamics only in terms of
Wigner-covariant relative variables, we have to quantize the
particle Wigner-covariant 3-variables $\eta^r_i$, $\kappa_{ir}$ and
then to define the physical Hilbert space by adding the quantum
version of the constraints a la Gupta-Bleuler.\medskip

The main implications for relativistic entanglement is that in
special relativity the zeroth postulate for composite systems does
not hold: Einstein's notion of separability is not valid since in $H
= H_{com, HJ} \otimes H_{rel}$ the composite system must be
described by means of relative variables in a Wigner 3-space (this
is a type of weak form of  relationism different from the
formulations connected to the Mach principle). Due to the problem of
clock synchronization  and to the structure of the Poincare' group,
special relativity introduces a {\it kinematical non-locality} and a
{\it kinematical spatial non-separability}, which reduce the
relevance of {\it quantum non-locality} in the study of the
foundational problems of quantum mechanics. The relativistic
formulation of problems like the relevance of decoherence \cite{23}
for the selection of preferred robust pointer bases and the
emergence of quasi-classical macroscopic objects from quantum
constituents will have to be done in terms of relative variables.
Moreover, the control  of Poincare' kinematics will force to
reformulate the experiments connected with Bell inequalities and
teleportation in terms of isolated systems containing: a) the
observers with their measuring apparatus (Alice and Bob as
macroscopic quasi-classical objects); b) the particles of the
protocol (but now the ray of light, the "photons" carrying the
polarization, move along null geodesics); c) the environment
(macroscopic either quantum or quasi-classical object).

The final challenge will be a consistent inclusion of the
gravitational field, at least at the post-Newtonian level!

\end{document}